\documentclass[11pt,reqno]{amsart}
\usepackage{amssymb}

\newcommand{\ndash}{\nobreakdash-\hspace{0pt}}
\newcommand{\Ndash}{\nobreakdash--}

\newcommand{\ii}{{\mathrm{i}}}
\newcommand{\dd}{{\mathrm{d}}} 
\newcommand{\ee}{{\mathrm{e}}}

\theoremstyle{remark}

\newtheorem*{Ack}{Acknowledgment}

\theoremstyle{definition}

\newcommand{\Poiss}[2]{\left\{{\,{#1}\,,\,{#2}\,}\right\}}

\newcommand{\vevo}[1]{{\left\langle\;{#1}\;\right\rangle}_0}

\newcommand{\bbR}{{\mathbb{R}}}

\newcommand{\de}{\partial}

\newcommand{\calC}{\mathcal{C}}

\newcommand{\calG}{\mathcal{G}}
\newcommand{\calO}{\mathcal{O}}

\newcommand{\Gfor}{{\mathcal{G}_\mathrm{form}}}
\newcommand{\bGfor}{{\Bar{\mathcal{G}}_\mathrm{form}}}

\newcommand{\frg}{{\mathfrak{g}}}

\DeclareMathOperator{\LL}{L}

\begin{document} 

\title{Poisson sigma models and deformation quantization}
%\date{}

\author[A.~S.~Cattaneo]{Alberto~S.~Cattaneo${}^*$}

\author[G.~Felder]{Giovanni Felder${}^{**}$}

\thanks{A.~S.~C. acknowledges partial support of SNF Grant
No.~2100-055536.98/1}

\maketitle
\centerline{\tiny $^*$Institut f\"ur Mathematik, Universit\"at Z\"urich,
CH-8057 Z\"urich, Switzerland. E-mail: \texttt{asc@math.unizh.ch}}  
\centerline{\tiny $^{**}$D-MATH, ETH-Zentrum, CH-8092 Z\"urich, Switzerland.
 E-mail: \texttt{felder@math.ethz.ch}}

\begin{abstract}
This is a review aimed at a physics audience on the relation between
Poisson sigma models on surfaces with boundary and deformation quantization. 
These models are topological open string theories. 
In the classical Hamiltonian approach, we describe the 
reduced phase space and its structures (symplectic
groupoid), explaining in particular
the classical origin of the
non-commutativity of the string end-point coordinates. We also review the
perturbative Lagrangian approach and its connection with Kontsevich's
star product. Finally we comment on the relation between the two 
approaches.
\end{abstract}
\section{Introduction}

A Poisson manifold is a smooth manifold endowed with a Poisson bivector
field, viz., a skew-symmetric contravariant tensor $\alpha$ of rank $2$
satisfying the Jacobi identity
\begin{equation}\label{J}
\alpha^{i[j}\de_i\alpha^{kl]}=0,
\end{equation}
where $[\cdots]$ denotes a sum over cyclic permutations of the indices
included in the square brackets. Using a Poisson bivector field
one can define the Poisson bracket $\Poiss fg:=\alpha^{ij}\de_if\de_jg$
of any two smooth functions $f$ and $g$. Two typical examples are $i$) a
symplectic manifold, and $ii$) $M=\frg^*$ with $\frg$ a Lie algebra and
$\alpha^{ij}(x)=f^{ij}_kx^k$, where $f^{ij}_k$ are the structure constants in
some basis.

The Poisson sigma model \cite{I,SS} is a sigma model whose worldsheet is
a connected, oriented, smooth $2$\ndash manifold (possibly with
boundary) $\Sigma$ and whose
target is a (say, $m$\ndash dimensional)
Poisson manifold $(M,\alpha)$. The fields of the model are a
map $X\colon \Sigma\to M$ together with a $1$\ndash form $\eta$,
with $\eta(u)$, $u\in\Sigma$, taking values in the cotangent space of
$M$ at $X(u)$. We will use Greek indices for worldsheet coordinates
and Latin indices for target coordinates. The action functional of the
Poisson sigma model reads then
\begin{equation}\label{S}
S(X,\eta) = \int_\Sigma \left[
\eta_{\mu i}\de_\nu X^i +
\frac12\alpha^{ij}(X)\eta_{\mu i}\eta_{\nu j}
\right]\dd u^\mu\dd u^\nu.
\end{equation}
If $\Sigma$ has a boundary, we choose the
boundary conditions $v^\mu(u)\eta_{\mu i}(u)=0$, $u\in\de\Sigma$, for any
vector $v$ tangent to the boundary.

In Section~\ref{cha} we review some results of \cite{CF2} about the classical
reduced phase space and its groupoid structure.
In Section~\ref{pla} we describe briefly the critical points of the action
functional \eqref{S}, its symmetries and its perturbative quantization and explain
its connection with star products.
For more details, we refer to \cite{CF1}.
Finally, we comment on the action of target diffeomorphisms and on the relation
between Kontsevich's star product and Weinstein's program based on symplectic groupoids.
\begin{Ack}
A. S. C. thanks the organizers of the meeting 
``BRANE NEW WORLD and Noncommutative Geometry,''
held in Turin in October 2000, for a kind
invitation to this most interesting conference.
\end{Ack}

\section{Classical Hamiltonian approach}\label{cha}
The action \eqref{S} describes the propagation of a topological open string.
Let us choose locally a time coordinate $t=u^0$ and denote by $u=u^1$ the
space coordinate. As we are describing an open string, $u$ will belong
to a closed interval $I$ which we may as well identify with $[0,1]$.
Then the Lagrangian function corresponding to \eqref{S} reads
\begin{equation}\label{L}
L(X,\zeta;\beta) = \int_I \left[
-\zeta_i\dot X^i + \beta_i\left(
{X^i}'+\alpha^{ij}(X)\zeta_j
\right)\right]\dd u,
\end{equation}
with $\beta_i=\eta_{0i}$, $\zeta_i=\eta_{1i}$, $\dot X=\de X/\de t$ and
$X'=\de X/\de u$.
The first term in the Lagrangian tells us that $\zeta$ and $X$ are 
canonically conjugated variables;\footnote{\label{f-mani}In order to make
the discussion mathematically precise, one should
actually introduce a structure of manifold on the space of fields. This
requires specifying which kind of maps one takes into consideration. If
one chooses the minimal conditions for \eqref{C} to make sense---i.e.,
$\zeta$ continuous and $X$ $1$\ndash differentiable---, one can then
give the space of fields the structure of an infinite-dimensional
symplectic manifold locally modeled on a Banach space; in this case,
most facts about finite-dimensional manifolds remain true.}
viz., we have the Poisson brackets
\begin{equation}\label{P}
\Poiss{\zeta_i(u)}{X^j(v)}=\delta_i^j\delta(u-v),
\end{equation}
while all other Poisson brackets vanish.
In the second term the new variable $\beta$ appears.
It has to be thought
of as a Lagrange multiplier imposing the constraints
\begin{equation}\label{C}
{X^i}'+\alpha^{ij}(X)\zeta_j=0.
\end{equation}
We will denote in the following by $\calC$ the space of solutions 
to \eqref{C}. Equivalently, we may define $\calC$ as the common zero set
of the functions
\begin{equation}\label{H}
H_\beta=\int_I \beta_i\left(
{X^i}'+\alpha^{ij}(X)\zeta_j
\right)\dd u,
\end{equation}
for all $\beta$ vanishing on the boundary of $I$ (since $\beta_i=\eta_{0i}$
and $\eta_{0i}$ is the component of $\eta$ tangent to the boundary).
It is now easy to check that the $H_\beta$s are first-class constraints:
viz., their Poisson brackets vanish on $\calC$, i.e.,
upon using \eqref{C}.
In fact, let
\[
\delta_\beta X^i(u):=\Poiss{H_\beta}{X^i(u)},\qquad
\delta_\beta \zeta_i(u) :=\Poiss{H_\beta}{\zeta_i(u)}
\]
denote the Hamiltonian vector field of $H_\beta$ applied to the fields
of the model: viz.,\footnote{\label{f-cov}The 
expression on the right hand side of
\eqref{dzeta} is clearly not covariant under target coordinate transformations. 
However, one can easily check
that the error
is proportional to the constraint \eqref{C}. So on $\calC$ the
symmetries are well-defined.}\begin{subequations}\label{d}
\begin{align}
\delta_\beta X^i &= -\alpha^{ij}(X)\beta_j,\label{dX}\\
\delta_\beta \zeta_i &= \beta_i'+\de_i\alpha^{rs}(X)\zeta_r
\beta_s.\label{dzeta}
\end{align}
\end{subequations}
Then we have,
\begin{multline*}
\delta_\beta\left(
{X^i}'+\alpha^{ij}(X)\zeta_j
\right)=
(-\alpha^{ij}(X)\beta_j)'-\de_l\alpha^{ij}(X)\alpha^{lr}(X)\beta_r\zeta_j+\\
+\alpha^{ij}(X)\beta_j'+\alpha^{ij}(X)\de_j\alpha^{rs}(X)\zeta_r\beta_s,
\end{multline*}
which vanishes upon using \eqref{C} and \eqref{J}. So we define the
reduced phase space $\calG$ as $\calC$ modulo the symmetries \eqref{d}.

\subsection{``Topological" nature of the model}\label{ss-top}
The absence of a Hamiltonian in \eqref{L} implies that the system has no 
dynamics, so it is invariant under diffeomorphisms in the time direction.
As for the space direction, consider an infinitesimal diffeomorphism of
the interval $I$, i.e., a vector field $\nu$ (which we may identify
with a function on $I$) vanishing at the boundary. Its action on the
field $X$ is given by the Lie derivative
\[
\LL_\nu X^i =\nu {X^i}'\stackrel{\text{on $\calC$}}=
-\nu\alpha^{ij}(X)\zeta_j.
\]
Thus, we can identify the action of $\nu$ with a symmetry \eqref{dX}
generated by $\beta_i=\nu\zeta_i$. Also observe that 
$\LL_\nu\zeta=(\nu\zeta)'=\beta'$ ($\zeta$ is a $1$\ndash form on $I$).
Since $\zeta_r\beta_s$ is symmetric in $r$ and $s$ for our choice of
$\beta$, it turns out that the action of $\nu$ on the pair of fields
$X$ and $\zeta$ corresponds to a symmetry transformation
\eqref{d}.\footnote{This construction actually works only if we consider
smooth maps $X$ and $\zeta$ which seems to be in contradiction with the
assumptions of footnote~\ref{f-mani}. The point is that one can show
that each class of solutions contains a smooth representative.}
As a result, the model is invariant under worldsheet diffeomorphisms.

We also observe that the actual number of degrees of freedom is finite.
Namely, one may show that the dimension of the tangent space of $\calG$
is finite and equal to $2m$, with $m=\dim M$. In fact,
let $(X,\zeta)$ be a representative of a class of solutions to \eqref{C}
modulo \eqref{d}.
Consider infinitesimal perturbations $\xi$ and
$\varsigma$ of $X$ and $\zeta$ respectively. The linearization of
\eqref{C} then reads
\[
{\xi^i}'+A_k^i\xi^k=-\alpha^{ij}(X)\varsigma_j,
\]
with $A_k^i(u)=\de_k\alpha^{ij}(X(u))\zeta_j(u)$.
This has a unique solution once $\xi^i(0)$ and $\varsigma$ are given:
\begin{equation}\label{xi}
\xi^i(u)=(V^{-1}(u))_k^i\left(\xi^k(0)-\int_0^u V_r^k(v)\alpha^{rs}(X(v))
\varsigma_s(v)\;\dd v\right),
\end{equation}
where $V$ is the path-ordered integral of $A$, i.e., the solution of
$V_i^j(u)'=V_k^j(u)A_i^k(u)$, $V_i^j(0)=\delta_i^j$. The symmetry
transformation \eqref{dX} does not change the boundary values of $X$ since
$\beta$ vanishes on the boundary; so the $m$ parameters $x^i:=\xi^i(0)$ are
well-defined. Now consider the linearization of the symmetry
\eqref{dzeta} on $\varsigma$:
$\delta_\beta\varsigma_i=\beta_i'-\beta_sA_i^s$.
It follows that $p_j:=\int_0^1\varsigma_i(u)(V^{-1}(u))^i_j\;\dd u$ is
another set of $m$ invariants.

Thus, every infinitesimal perturbation $(\xi,\varsigma)$ determines
uniquely $2m$ parameters $(x,p)$. Conversely, given $(x,p)$, we can
first choose $\varsigma_i(u)=p_j (K^{-1})^j_i$, with
$K^i_j=\int_0^1 (V^{-1}(u))^i_j\;\dd u$, 
and then compute $\xi$ using \eqref{xi}.

\subsection{``Noncommutativity" of the end-points}
Since $\beta$ vanishes at the boundary, the boundary values of $X$ are
invariant under \eqref{dX}. We set $x^i=X(0)$ and $y^i=X(1)$.
We want to show that, using the Poisson bracket induced from \eqref{P}
on $\calG$, we get
\begin{equation}\label{xx}
\Poiss{x^i}{x^j}=\alpha^{ij}(x),\qquad
\Poiss{y^i}{y^j}=-\alpha^{ij}(y),\qquad
\Poiss{x^i}{y^j}=0.
\end{equation}
The main problem in this computation is that the functions $X^i(0)$ and
$X^i(1)$ do not possess Hamiltonian vector fields. However, upon using
\eqref{C}, we can equivalently define the ``regularized" versions
\begin{align*}
x^i&=X^i(u)+\int_0^u\alpha^{ij}(X(v))\zeta_j(v)\;\dd v,\\
y^i&=X^i(u)-\int_u^1\alpha^{ij}(X(v))\zeta_j(v)\;\dd v,
\end{align*}
for any $u\in(0,1)$. So we can compute
\[
\Poiss{x^i}{x^j}=
\Poiss{X^i(u)+\int_0^u\alpha^{ij}(X(v))\zeta_j(v)\;\dd v}{X^j(0)}=
\alpha^{ij}(X(0)),
\]
and proceed similarly for the other identities in \eqref{xx}.
(Observe that there is no need to regularize the second entry in the
Poisson bracket as it is enough that the first entry admits a
Hamiltonian vector field, and then one applies this field to the second
entry). 

\subsection{Composing solutions}\label{s-comp}
If we are given two solutions $(X_1,\zeta_1)$ and $(X_2,\zeta_2)$ to
\eqref{C} such that $X_1(1)=X_2(0)$, we may compose them into a new
solution $(X_3,\zeta_3)$:
\begin{align*}
X_3(u) &= \begin{cases}
X_1(2u) & 0\le u <\frac12,\\
X_2(2u-1) & \frac12\le u \le 1,
\end{cases}\\
\zeta_3(u) &= \begin{cases}
2\zeta_1(2u) & 0\le u <\frac12,\\
2\zeta_2(2u-1) & \frac12\le u \le 1.
\end{cases}
\end{align*}
To avoid a singularity at $u=1/2$, we should assume that $\zeta$ vanishes
at the boundary points;\footnote{This makes $\zeta_3$
continuous and $X_3$ $1$\ndash differentiable as required by
footnote~\ref{f-mani}.}
one can indeed prove that each class
$[(X,\zeta)]\in\calG$ contains a representative with this property.

This way we have constructed a partially defined product $\bullet$
on $\calG$; viz.,
we have a map $\bullet\colon\calG\times_M\calG\to\calG$, where
$\calG\times_M\calG$ is the space of pairs of classes of solutions with the
end-point of the first coinciding with the starting point of the second.
By the discussion of subsection~\ref{ss-top}, we obtain immediately that
this product is associative; in fact, the two possible ways of
combining three solutions are related by a reparametrization of the
interval $[0,1]$, which---as we have seen---can be obtained by a
symmetry transformation. Moreover, for each point $x$ of $M$, we have a
very peculiar solution: viz., $X(u)\equiv x$ and $\zeta(u)\equiv0$; this
solution plays the role of a unit at $x$ for the product $\bullet$.
Finally, for every solution $(X,\zeta)$, we have an inverse solution
$(\Bar X, \Bar\zeta)$, with $\Bar X(u)=X(1-u)$, $\Bar\zeta(u)=-\zeta(1-u)$.

The existence of all the structures described in the previous paragraph
is expressed by saying that $\calG$ is a {\sf groupoid} for $M$.
Actually, in the lucky cases when $\calG$ turns out to be a smooth
symplectic manifold, one can further prove that $\calG$ is a {\sf symplectic groupoid}
for $M$. This notion was introduced by   Karasev \cite{Ka}, Weinstein
\cite{W} and Zakrzewski
\cite{Z}. 
It means that some extra conditions are satisfied; the main ones
are the identities displayed in \eqref{xx} and
the fact that
$\{(a,b,c)\in\calG\times_M\calG\times\calG : a\bullet b=c\}$
is a Lagrangian submanifold of
$\calG\times\calG\times\Bar\calG$, where $\Bar\calG$ as a manifold is the
same as $\calG$ but with opposite symplectic structure.

\subsection{The formal symplectic groupoid}
Locally one can use \eqref{dzeta} to set $\zeta$ constant. If one denotes by
$p$ this constant, then a point of the reduced phase space may be taken
as a solution of
\begin{equation}\label{fC}
{X^i}'(u)+\alpha^{ij}(X(u))p_j=0.
\end{equation}
This set of
equations defines globally a space $\calG'$ that coincides only locally
with our $\calG$. This space $\calG'$ was studied in \cite{Ka,KM} and
called the {\sf local symplectic groupoid}: it
always admits the structure of a smooth symplectic manifold (globally, despite
of the name) but has an
associative product that is defined only in a neighborhood of the
constant solutions. This is the price to pay with respect to our
$\calG$, which has a globally defined groupoid structure (but may fail
to be globally a manifold).

It is possible however to take an intermediate point of view and get a
globally defined groupoid structure compatible with a globally defined
smooth symplectic structure but in a formal sense. Namely, we may
consider solutions to \eqref{C} modulo symmetries \eqref{d} as formal
power series in $\alpha$ and its derivatives. We will denote this {\sf formal
symplectic groupoid} by $\Gfor$. Let us describe briefly its construction.
As a manifold $\Gfor$ is just $T^*M$. In fact given a point $(x,p)\in
T^*M$, we associate to it the pair $(X,\zeta)$, where $\zeta\equiv p$
and $X$ is the formal solution
of \eqref{fC} with initial
condition $X(0)=x$. Conversely, given a solution to \eqref{fC}, it is
possible to construct a formal path of symmetries \eqref{d} that makes
$\zeta$ constant. Then we set $x=X(0)$ and $p=\zeta$.\footnote{Notice
however that the expression of $p$ in terms of the original nonconstant
$\zeta$ will be rather complicated, though in principle iteratively
computable. It will involve a formal series of iterated integrals of
$\zeta$ and of $\alpha$ and its derivatives, giving a sort of
generalization of the analogue expansion of the holonomy in the case
when $M$ is the dual of a Lie algebra and $\zeta$ can be viewed as a
connection.}
The symplectic structure can be easily discussed
by changing coordinates $(x,p)\mapsto(y,p)$ with $y=\int_0^1 X(u)\;\dd
u=x+O(\alpha)$ where $X$ is the solution corresponding to $(x,p)$.
These are Darboux coordinates; in fact, the symplectic structure on the space
of maps corresponding to \eqref{P} is the differential of
$\int_0^1\zeta_i(u)\delta X^i(u)\;\dd u$ (where $\delta$ denotes the
differential on the space of fields). Restricting this one form to $\Gfor$
one gets $p_i\dd y^i$, whose differential is the symplectic form
$\dd p_i\dd y^i$. Repeating verbatim the construction of
subsection~\ref{s-comp}, we can define a product on $\Gfor$
which will be denoted by the same symbol $\bullet$.

Let us now consider the case when $M$ is a domain in $\bbR^m$. Then
$\Gfor$ is $M\times\bbR^m\ni(y,p)$, and we have two globally defined
projections. 
We see then $\Gfor\times\Gfor\times\bGfor$ as a trivial bundle over
$\bbR^m\times\bbR^m\times M$. The fact that the groupoid product defines a
Lagrangian submanifold of $\Gfor\times\Gfor\times\bGfor$ can now be
expressed in terms of a generating function; viz.,
there exists a function $F(p_1,p_2,y_3)$ on
$\bbR^m\times\bbR^m\times M$ such that
\begin{equation}\label{F}
\left(\frac{\de F}{\de p_1},p_1\right) \bullet
\left(\frac{\de F}{\de p_2},p_2\right) =
\left(y_3,\frac{\de F}{\de y_3}\right).
\end{equation}
Following an idea of Weinstein \cite{W2},\footnote{In terms of geometric
quantization, we consider a trivial line bundle over
$\Gfor\times\Gfor\times\bGfor$ with connection
$A=(-y_1\cdot\dd p_1-y_2\cdot\dd p_2-p_3\cdot\dd y_3)/\hbar$. Then
$\ee^{\frac\ii\hbar F(p_1,p_2,y_3)}$ is covariantly constant when
restricted to the Lagrangian submanifold determined by $F$. Moreover, it
is covariantly constant along the polarization generated by
$\de/\de y_1$, $\de/\de y_2$ and $\de/\de p_3$.
On the other hand, if we choose $A=-y\cdot\dd p/\hbar$ as a connection on
$\Gfor$, then $f(y)\ee^{\frac\ii\hbar p\cdot y}$ is covariantly constant
along the polarization $\de/\de p$ (which is transverse to the above
$\de/\de y$).}
we can define a ``semiclassical'' product on
$C^\infty(M)[[\hbar]]$ by the formula
\begin{equation}\label{starcl}
f\star_\mathrm{sc}g(y)=
\int \ee^{-\frac\ii\hbar F(p_1,p_2,y)}
f(y_1)\ee^{\frac\ii\hbar p_1\cdot y_1}
g(y_2)\ee^{\frac\ii\hbar p_2\cdot y_2}
\frac{\dd^m y_1\dd^m p_1\,\dd^my_2\dd^mp_2}{(2\pi\hbar)^{2m}},
\end{equation}
with $p\cdot y:=p_iy^i$. 
An explicit computation shows that
$F(\hbar p_1,\hbar p_2,y_3)=\hbar y_3^i(p_1+p_2)_i
+\frac{\hbar^2}2 \alpha^{ij}(y_3)(p_1)_i(p_2)_j +O(\hbar^3)$, so that
$f\star_\mathrm{sc}g=fg+\frac{\ii\hbar}2\alpha^{ij}\de_if\de_jg
+O(\hbar^2)$. Moreover, one can check that $F(p_1,0,y_3)=y_3\cdot p_2$ and
that $F(0,p_2,y_3)=y_3\cdot p_2$ which implies
$f\star_\mathrm{sc}1=1\star_\mathrm{sc}f=f$. Finally, the structure
of $F$ implies that 
$f\star_\mathrm{sc} g=\sum_{n=0}^\infty \hbar^n B_n^\mathrm{sc}(f,g)$ where the
$B_n^\mathrm{sc}$s are differential operators w.r.t.\ both entries.

Unfortunately, although $\star_\mathrm{sc}$ is defined using the
associative groupoid product $\bullet$,
nothing guarantees that it is associative. This happens in some lucky
instances; e.g., in the case of the dual of a Lie algebra, the above
construction yields $F(p_1,p_2,y_3)=y_3\cdot\mathrm{CBH}(p_1,p_2)$,
where $\mathrm{CBH}$ is the Campbell--Baker--Hausdorff formula.
In general, we may expect that \eqref{starcl} is only an approximation
to an associative product obtained by adding to $F$ corrections in
$\hbar$, as we will see in the next Section.

\section{Perturbative Lagrangian approach}\label{pla}
We now turn back to the action functional \eqref{S}.
Its critical points
are solutions to the equations
\begin{gather*}
\de_\nu X^i+\alpha^{ij}(X)\eta_{\nu j}=0,\\
\epsilon^{\mu\nu}\left(\de_\mu\eta_{\nu i}+ 
\frac12\de_i\alpha^{jk}(X)\eta_{\mu j}\eta_{\nu k}\right)=0,
\end{gather*}
where $\epsilon^{\mu\nu}$ is the totally antisymmetric Levi-Civita tensor.
There are particular critical points, 
which we will call {\sf trivial},
consisting of a constant map $X\equiv x\in M$ together
with $\eta\equiv0$.

The symmetries of the action functional can be deduced from the symmetries
\eqref{d} obtained in the Hamiltonian formalism. More precisely,
we can introduce a ghost $\beta$ (with $\beta(u)$ in the cotangent
space of $M$ at $X(u)$ and $\beta\equiv0$ on the boundary) and the
BRST operator $\delta$ acting as follows:
\begin{subequations}\label{dd}
\begin{align}
\delta X^i &= -\alpha^{ij}(X)\beta_j,\label{ddX}\\
\delta\eta_{\mu i} &= \de_\mu\beta_i+\de_i\alpha^{rs}(X)\eta_{\mu r}
\beta_s.\label{ddeta}\\
\intertext{The BRST variations of $\beta$ may be taken to be}
\delta\beta_i &= \frac12 \de_i\alpha^{jk}(X)\beta_j\beta_k,\label{ddbeta}
\end{align}
\end{subequations}
so that $\delta$ squares to zero on shell. However, apart from some simple 
cases (e.g., when $M$ is the dual of a Lie algebra), this BRST operator does 
not square to zero off shell. So one has to use the BV formalism instead.
A simple way of applying it to this case is to use the AKSZ formalism 
\cite{AKSZ} as explained in \cite{P,CF3}. We will not enter into details 
here. Only observe that the AKSZ method provides
a solution to the classical master equation, but one has then to check that the
quantum master equation is also satisfied. This is
done in \cite{CF1} by using a regularization that breaks the rotational symmetry
of the disk. If however $\partial_i\alpha^{ij}=0$ in given coordinates, this
regularization is not necessary, and one obtains a cyclic product: 
$\int (f\star g)h\;\dd^mx=\int (g\star h)f\;\dd^mx$, see \cite{FS}.

The main observation now is that the boundary condition on $\beta$ together
with \eqref{ddX} implies that for any function $f$ on $M$ we may define a 
BRST invariant observable $f(X(u))$ where $u$ is any point on the boundary
of $\Sigma$.\footnote{More generally, one can consider observables
of the same form but with $u$ in the interior of $\Sigma$ if $f$ is a
Casimir function, i.e., $\alpha^{ij}\de_jf\equiv0$.}
Let us consider the simplest topology, i.e., assume that
$\Sigma$ is a disk. After picking three points $0$, $1$ and $\infty$ on its
boundary, we may define a product on $C^\infty(M)[[\hbar]]$ by
\begin{equation}\label{star}
f\star g(x)=\vevo{f(X(1))\,g(X(0))\,
\delta_x(X(\infty))},
\end{equation}
where $\delta_x$ is the delta function peaked at $x\in M$ and $\vevo{\ }$
denotes the (perturbative) expectation value w.r.t.\ the action \eqref{S}
around the trivial critical point.\footnote{Taking into account
all critical points may yield restrictions to the possible values
$\hbar$ can assume.}
It is not difficult to check that
$f\star1=1\star f=f$, that
$f\star g=fg+\frac{\ii\hbar}2\alpha^{ij}\de_if\de_jg+O(\hbar^2)$
and that in general $f\star g=\sum_{n=0}^\infty \hbar^n B_n(f,g)$ where the
$B_n$s are differential operators w.r.t.\ both entries ({\sf bidifferential
operators}).
Finally, the topological nature of the model allows us to move points on the boundary,
and this leads to the associativity of $\star$.
Thus, \eqref{star} defines a so-called {\sf star product} \cite{BFFLS}.

In \cite{CF1} this perturbative expansion has been studied mapping
the disk to the upper half plane (the marked point $\infty$ being mapped
to infinity) with the gauge fixing
$g^{\mu\nu}\de_\mu\eta_{\nu i}=0$, where $g^{\mu\nu}$ is the Euclidean
metric. It turns out to coincide with Kontsevich's formula \cite{Ko} of which
it provides a field-theoretical motivation.
Let us briefly describe the Feynman diagrams of this expansion.
A diagram of order $n$ (i.e., producing a coefficient $\hbar^n$) contains
$n$ vertices in the upper half plane 
and two vertices at the points $0$ and $1$ on the boundary: 
we call the former {\sf internal vertices} and the latter
{\sf external vertices}. From each internal vertex there depart two ordered,
oriented edges.
The head of an edge can then land on another internal
vertex ({\sf internal edge}) or an external vertex 
({\sf external edge}).\footnote{Tadpoles (i.e., edges that land on their
own starting point) would yield singular contributions and have to be removed,
as usual, by renormalization. Again this is not necessary if
$\partial_i\alpha^{ij}=0$ in given coordinates.} Each 
diagram then gives a bidifferential
operator by putting an $\alpha$ on each internal vertex, putting
the functions $f$ and $g$ on the external ones, and associating to each edge
a partial derivative acting on the object it lands on and with its index
contracted with the corresponding
index of the $\alpha$ placed at its starting point.
The bidifferential operator is then multiplied by a weight obtained
by multiplying propagators corresponding to edges and integrating
over the configuration space of all vertices. For more details, we refer
to \cite{Ko,CF1}.

Observe that the perturbative expansion requires to Taylor expand
the Poisson bivector field $\alpha$ and the functions $f$ and $g$, and
this in turn requires choosing local coordinates on $M$.
Changing coordinates is then rather involved. In particular, if $\xi$ is
a vector field on $M$, its usual action (Lie derivative) on functions is not
compatible with the star product (i.e., it is not a derivation). Its correct
deformation (see \cite{CF3}) is given by the formula
\[
A_\xi(f)(x)=\vevo{ f(X(1))\,\calO_\xi\,
\delta_x(X(\infty))}=\xi^i(x)\de_if(x)+O(\hbar),
\]
where $\calO_\xi$ is the BV observable whose classical part is
$\int_{\gamma}\eta_{\mu i}(u)\xi^i(X(u))\;\dd u$, where $\gamma$ is a path
that separates the disk into a component containing $1$ and a component
containing $\infty$. This deformed Lie derivative is a component of
Kontsevich's quasiisomorphim \cite{Ko}, and using its properties one
understands how to globalize the star product \cite{Ko,CFT}.

We now want to discuss the relations between Kontsevich's 
star product \eqref{star} and its semiclassical (possibly nonassociative)
counterpart \eqref{starcl}. Assume $M$ to be a domain in $\bbR^m$
and let $p\cdot y$ denote $p_iy^i$ for $y\in M$ and $p\in T^*_yM\simeq\bbR^m$.
Define $\phi_p(y)=\exp(-\ii p\cdot y/\hbar)$ and set
\begin{equation}\label{HF}
\ee^{-\frac\ii\hbar \Hat F(p_1,p_2,y;\hbar)}:=
\phi_{p_1}\star\phi_{p_2}(y).
\end{equation}
Then, by linearity, we can write
\begin{equation}\label{starcorr}
f\star g(y)=
\int \ee^{-\frac\ii\hbar \Hat F(p_1,p_2,y;\hbar)}
f(y_1)\ee^{\frac\ii\hbar p_1\cdot y_1}
g(y_2)\ee^{\frac\ii\hbar p_2\cdot y_2}
\frac{\dd^m y_1\dd^m p_1\,\dd^my_2\dd^mp_2}{(2\pi\hbar)^{2m}},
\end{equation}
which is the correct (associative) generalization of \eqref{starcl}.
We observe that $\Hat F$ has three interesting properties: $i$) it is
a formal power series in $\hbar$ (i.e., no negative powers); $ii)$ the
function $F:=\Hat F|_{\hbar=0}$ is a generating function of a symplectic
groupoid on $T^*M$ with canonical symplectic structure---i.e., $F$
satisfies \eqref{F}---; $iii)$ the coefficients
of the strictly positive powers in $\hbar$ correspond to Feynman diagrams
containing internal loops (i.e., loops consisting only of internal edges).

The first and the third properties follow from the fact that a Feynman diagram
contributing to $\phi_{p_1}\star\phi_{p_2}$ will carry a coefficient
$\hbar^{n-e}$ if it has $n$ internal vertices and $e$ external edges.
Observe moreover 
that 
$\Hat F$ is defined in terms of only those Feynman diagrams that are
still connected after deleting all external legs. 
This means that if the diagram has $n$ vertices, it can have at most $n+1$
external legs, and this proves $i)$. To prove $iii)$ observe that
strictly positive orders in $\Hat F$ correspond to diagrams whose number
of external legs is at most equal to the number of internal vertices; but
this means that the number of internal edges is at least equal to the number
of internal vertices, so that there must be an internal loop.
Property $ii)$, viz.\ equation \eqref{F}, is instead obtained 
by computing in saddle-point approximation
the products $(f\star g)\star h$ and $f\star(g\star h)$ which must be
equal thanks to associativity.

\section{Discussion}
In this note we have reviewed some early results of ours, in particular
\cite{CF1} and \cite{CF2}, and described some related ideas.
In Section~\ref{cha}, after a brief description of the structures
of the classical
reduced phase space of the Poisson sigma model found in \cite{CF2},
we have discussed how to apply an old idea of Weinstein's \cite{W2} to define
a star product, see \eqref{starcl}. 
At this level however there is no clue that the product
defined this way is associative. In Section~\ref{pla}, after recalling
the results of \cite{CF1} about the perturbative quantization of the Poisson
sigma model and its relation with Kontsevich's formula \cite{Ko},
we have seen how to obtain in this context a correction to Weinstein's
formula, see \eqref{starcorr}.

The natural question at this point is that if it had been possible to find
directly the corrections to \eqref{starcl} to make it associative.
A related idea stems from the observation that \eqref{starcl} looks like
an expectation value of a lattice formulation of the Poisson sigma model
with a particular gauge fixing. Namely, let us consider a lattice of three
triangular plaquettes. To each plaquette we associate a point $y$ of the domain
$M$ and to each
edge not on the boundary we associate an element $p$ of $\bbR^m$.
We take then $-F$ as the plaquette action. Formula \eqref{starcl}
is then obtained by setting the variable of one plaquette equal to the
given $y$ and by placing functions $f$ and $g$ in the two remaining plaquettes
and setting the $p$\ndash variable corresponding to the edge between the latter
two plaquettes equal to zero (``gauge fixing''). 
Can this be reformulated by constructing a lattice BV action for the Poisson
sigma model corresponding to the above description? It may then happen that
this action does not satisfy the quantum master equation and that its
iterative solution explains the higher order terms in $\Hat F$.
Another possibility is that, on such a small lattice, we cannot obtain
associativity which might however be recovered by taking a bigger one.
How big should it be? Does it have to be infinite as long as $\alpha$ is not
linear, or can one expect to be able to adjust the lattice size according
to degree of $\alpha$ in the given coordinates?

Finally, the definition of $\Hat F$ in \eqref{HF} has a nice interpretation
in terms of quantum field theory; viz., we may absorb the terms 
$-p_1X(0)$ and $-p_2X(1)$ in the action \eqref{S}. These singular terms
can be regularized by removing a small neighborhood of $0$ and $1$ from
the integration domain of the action and adjusting
the boundary conditions of $\eta$ consequently. The result is an expansion
not around a trivial solution but around a solution that maps the negative
real axis, the interval $(0,1)$ and the interval $(1,\infty)$ to the
three points $y_1$, $y_2$ and $y$, respectively, obeying the formula
$(y_1,p_1)\bullet (y_2,p_2)=(y,p)$ (for a uniquely determined $p$).

\end{document}